# Using Floating Gate Memory to Train Ideal Accuracy Neural Networks

Sapan Agarwal, *Member, IEEE*, Diana Garland, John Niroula, Robin B. Jacobs-Gedrim, *Member, IEEE*, Alex Hsia, Michael S. Van Heukelom, Elliot Fuller, Bruce Draper, Matthew J. Marinella, *Senior Member, IEEE*

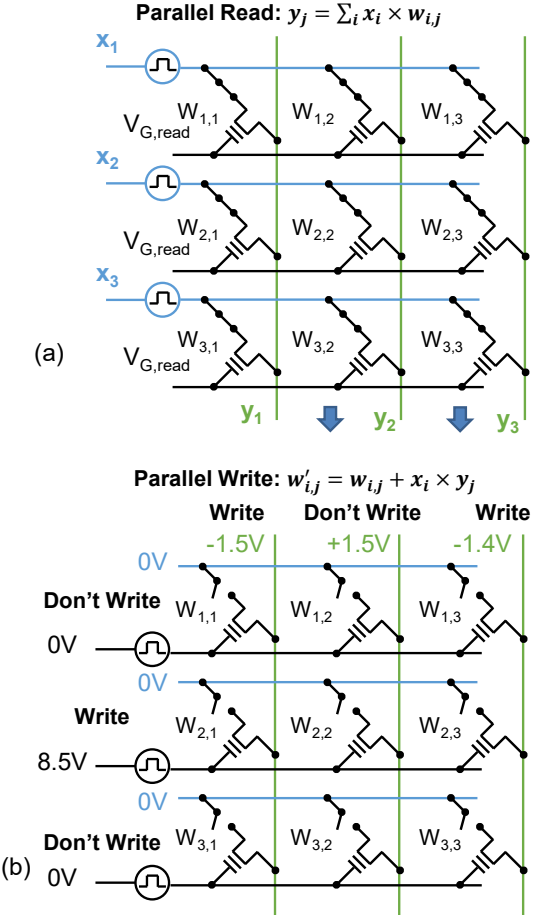

Fig. 1: (a) A vector matrix multiply is illustrated. A fixed read voltage is applied to all gate terminals. Access transistors (drawn as a switch) are biased on. Pulses of varying lengths are applied along the rows, and the resulting current is integrated along the columns. The transpose matrix vector multiply can be performed by applying pulses to the columns and reading along the rows. (b) A parallel write, or outer product update, for a 10V write is shown with the corresponding biases labeled. The access transistors are open circuited. Selected devices see up to the full 10V across $V_{GS}$ while unselected devices see a maximum of 7V across $V_{GS}$. The last column has a write voltage of -1.4V, applying 9.9V across $W_{2,3}$ resulting in smaller state change than a full 10V write. The amount written can be controlled by varying the voltage or pulse length.

*Abstract*— Floating gate SONOS (Silicon-Oxygen-Nitrogen-Oxygen-Silicon) transistors can be used to train neural networks to ideal accuracies that match those of floating point digital weights on the MNIST dataset when using multiple devices to represent a weight or within 1% of ideal accuracy when using a single device. This is enabled by operating devices in the subthreshold regime, where they exhibit symmetric write nonlinearities. A neural training accelerator core based on SONOS with a single device per weight would increase energy efficiency by 120X, operate 2.1X faster and require 5X lower area than an optimized SRAM based ASIC.

## I. Introduction

Analog accelerators promise to improve the energy and latency of training a neural network (NN) by more than a 100X over an optimized ASIC[1]. Analog matrix operations are used to process each memory element in parallel and thereby eliminate data movement as illustrated in Fig 1 [2]. However, this requires devices with high resistance, low write variability and low write nonlinearity[3]. Resistive memory devices have been used to represent synaptic weights, but the write variability and asymmetric write nonlinearity in current resistive memory device technology prevents the weights from being learned to high accuracy[3, 4]. Algorithmic and circuit techniques help improve accuracy [5, 6], but neural network accuracy is not ideal. Novel lithium [7] and polymer [8] based devices with excellent analog properties have been demonstrated, but will require continued work to integrate into modern CMOS foundries. In this paper, we show that a conventional floating gate memory, commonly available in foundries, can be used train a neural network to within 1% of that achieved with floating point weights on MNIST dataset (ideal accuracy). It has been shown that floating gate memories can be used to create accurate inference accelerators[9, 10]. We extend this to online training. Furthermore, the recently

Submitted Dec 31st, 2018. This work was supported by Sandia National Laboratories' Laboratory Directed Research and Development Program. Sandia National Laboratories is a multimission laboratory managed and operated by National Technology and Engineering Solutions of Sandia LLC, a wholly owned subsidiary of Honeywell International Inc. for the U.S. Department of Energy's National Nuclear Security Administration under contract DE-NA0003525.

S. Agarwal (email: sagarwa@sandia.gov) and E. Fuller are with Sandia National Laboratories, Livermore, CA
D. Garland, J. Niroula, R. B., Jacobs-Gedrim, A. Hsia, M. S. Van Heukelom, B. Draper, and M. Marinella (email: mmarine@sandia.gov) are with Sandia National Laboratories, Albuquerque, NM demonstrated periodic carry technique with multiple cells per weight [5] enables training to ideal accuracy. We also estimate that an 8-bit floating gate based accelerator will have training energy, latency, and area advantages of 120X, 2.1X, and 5X respectively versus performing the same training tasks with an optimized SRAM-based ASIC.

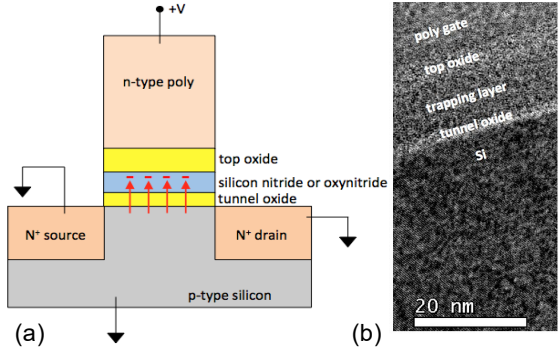

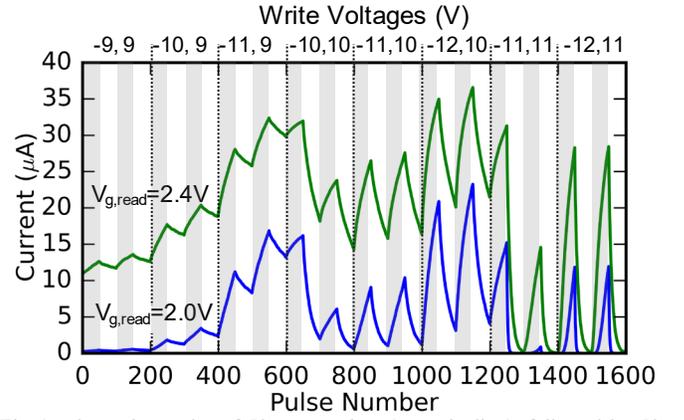

Fig 2: (a) A SONOS memory is schematically illustrated (b) A Transmission electron micrograph (TEM) of the gate stack is shown. The channel length of the device is 1.2 μm and the channel width is 7 μm. The ONO layer was grown in a tunnel oxidation furnace (VTR-20) in a dilute nitrous oxide ($N_2O$) atmosphere at 750°C.

Fig 4: Alternating series of 50 erase pulses (gray shading), followed by 50 program pulses (white shading) are applied for different write voltages. $V_S=V_B=0V$ and $V_D$ is floating. After applying a write pulse, the conductance is measured at $V_{GS}=2V$ and 2.4V and $V_{DS}=0.1V$. Write voltages of $V_{GS}=-11V$, 10V give a reasonable on/off range and high write linearity. Increasing the erase voltage to -14V broke the device.

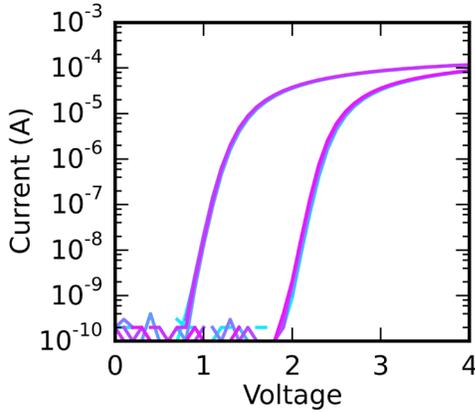

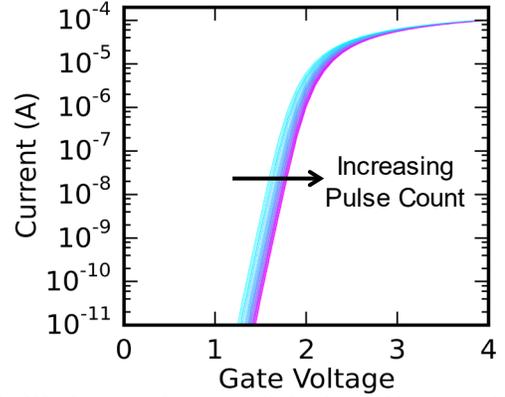

Fig 3: The binary memory window of the SONOS cell is shown. Alternating -11V, 2.5 ms, erase pulses and 10V, 2.5 ms, program pulses are applied. The pulse lengths can be increased to further increase the memory window.

Fig 5: 50 10V, 10 μs set pulses are applied and an I-V is measured after each pulse. During the analog write the threshold only shifts by about 200 mV, instead of the full 1-2V of a memory write.

In order to accelerate NN training using backpropagation, three kernels need to be accelerated: vector matrix multiplication (VMM), matrix vector multiplication (MVM) and Outer Product Update (OPU) [2], as shown in Fig 1. To accelerate both VMM and MVM, the source needs to be connected to the rows and the drain connected to the columns (or vice versa). During the OPU (parallel write), this configuration requires an access transistor for each memory cell to disconnect the drain from the rows. The access transistor prevents hot electron injection and junction breakdown. It also prevents large currents from flowing between the source and drain, which would cause unacceptable energy consumption and parasitic voltage drops in an array.

## II. DEVICE CHARACTERIZATION

The SONOS (Silicon-Oxygen-Nitrogen-Oxygen-Silicon) memory cell illustrated in Fig 2 was fabricated and characterized. The binary memory operation is illustrated in Fig 3. A reasonable 1V memory window is shown. Using longer write pulses or higher voltages can give a larger memory window. In Fig 4, we characterize the analog properties of the device for different write voltages. The write voltage used determines the number of analog states and write linearity. Write pulses of $V_{GS}=-11V$ for 10μs and $V_{GS}=+10V$ for 10 μs were chosen as the lowest voltages that give a reasonable $G_{high}/G_{low}$ ratio and high linearity in the conductance versus pulse characteristic. The threshold shift during the analog write is illustrated in Fig 5 and is only about 200mV. This is because only a ~10X $G_{high}/G_{low}$ ratio is needed for analog operation.

To analyze the effect of drain bias while programming the cell in an array, we investigated the effects of different source-drain configurations, including, $V_{DS}=0$, $V_{DS}=3V$, and floating/High-Z (Fig 6). Ideally, $V_{DS}=0$ during write. To achieve a condition close to this, an access transistor is used to float the drain, resulting in the drain floating condition. To see what would happen without an access transistor, VDS=±3 was also applied across the drain. Fortunately, changing $V_{DS}$ does not significantly affect the state written as both the source and body are grounded. This indicates that there is potential for writing without an access transistor to float the drain.

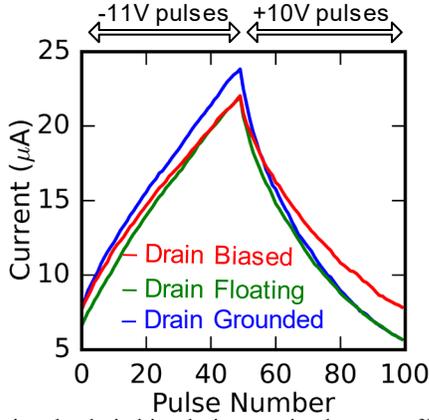

Fig 6: Changing the drain bias during a write does not affect the write properties. In all cases, $V_{body}$=0V. The biased case corresponds to the conditions without an access device: Vs=0, Vd=3V during program, -3V during erase.

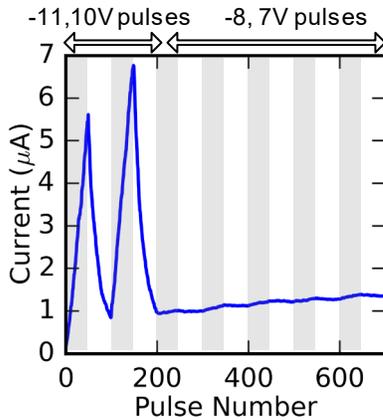

Fig 7: Alternating series of erase (gray shading) and program (white shading) pulses are applied. Lowering the write voltage from $V_{GS}$= -11, 10V to -8, 7V inhibits significant state change.

Nevertheless, we model an access device in subsequent area projections to eliminate parasitic currents during a write and to improve reliability by preventing hot electron injection. Eliminating the transistor would require redesigning the floating gate cell to limit the on-state current to limit the parasitic currents during a write.

It has also been verified that unselected devices do not change state under partial gate-bias conditions, with $V_{GS}$= -8V for erase and $V_{GS}$= +7V for program as illustrated in Fig 7. The access transistor only must block half the difference between the selected and unselected write voltages, reducing the size requirement of this transistor. If the write voltage is $V_{GS}$=10V and the unselected write voltage is 7V, the access transistor will have to hold off 1.5V.

The key limitation in neural network training accuracy is the asymmetric nonlinearities during a write [3]. With an asymmetric nonlinearity, alternating program and erase pulses that can occur at the end of training cause the weight to decay to a midpoint value. Nevertheless, neural networks can train to high accuracy with symmetric write nonlinearities [3]. To optimize the write nonlinearity, the gate read voltage needs to be optimized as shown in Fig 8. Choosing the correct read gate voltage will have a dramatic impact on the neural network work accuracy. As $V_{G,read}$ is lowered from 2.6V to 1.4V, the nonlinearity changes from an asymmetric nonlinearity to a symmetric linearity. By lowering $V_{G,read}$ the device is operating in the subthreshold regime. In this regime the magnitude of the change in conductance after a write pulse primarily depends on the starting state and not the sign of the write voltage. Achieving a symmetric nonlinearity is critical to enabling high accuracy training of neural networks.

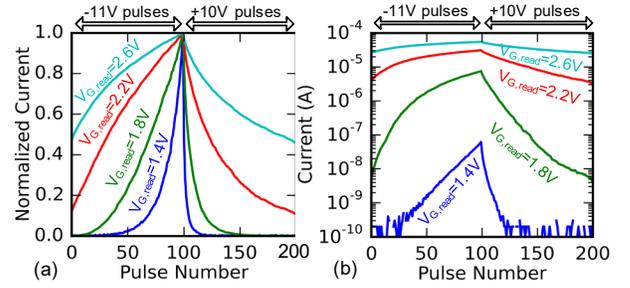

Fig 8: 100 erase pulses ($V_{GS}$= -11V, 10μs) followed by 100 program pulses ($V_{GS}$= 10V, 10μs) are applied and the current is measured at different read gate voltages. (a) The normalized current and (b) absolute value of the current is shown. Decreasing $V_{G,read}$ significantly reduces the write nonlinearity and changes the nonlinearity from an asymmetric nonlinearity to a symmetric nonlinearity.

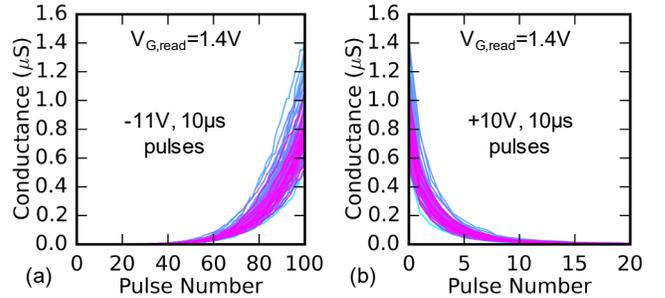

Fig 9: Alternating series of 100 erase pulses followed by 100 program pulses are applied. The conductance after each pulse is read at $V_{DS}$ = 100 mV and the measurement is repeated 50 times to collect statistics.

To characterize the analog statistics, a series of increasing and decreasing pulses were applied as illustrated in Figs 9-11. The conductance versus pulse number is plotted in Fig 9. In Fig 10, the conductance change at $V_{G,read}$=1.4V for different starting conductances is extracted from the pulsing data in Fig 9. We see the symmetric write nonlinearity where the conductance change is directly proportional to the starting state. In Fig 11(a) at $V_{G,read}$=2.6V, this reverses resulting in an asymmetric nonlinearity. The asymmetric nonlinearity results in significantly lower training accuracies.

A remaining challenge is understanding analog endurance in a floating gate device. A typical analog write pulse is only 0.1% or less of the length of a digital memory pulse[3], potentially increasing the endurance by three orders of magnitude or more. Furthermore, neural network training is also resilient to occasional device failure[4]. If needed, it's also possible to tradeoff retention for endurance.

### III. NEURAL NETWORK SIMULATION

To simulate the accuracy of a neural network based on this

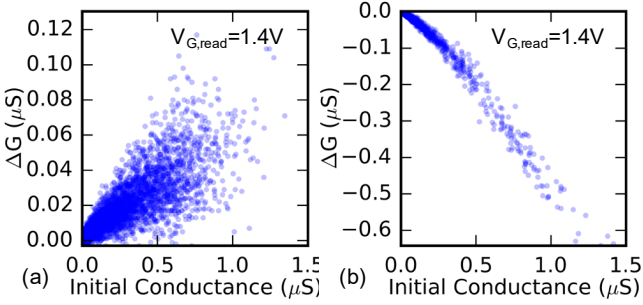

Fig 10: At $V_{G,read}$=1.4V, the conductance change is symmetric between program and erase (small changes at a low initial state and large changes at a high initial state) leading to a high training accuracy.

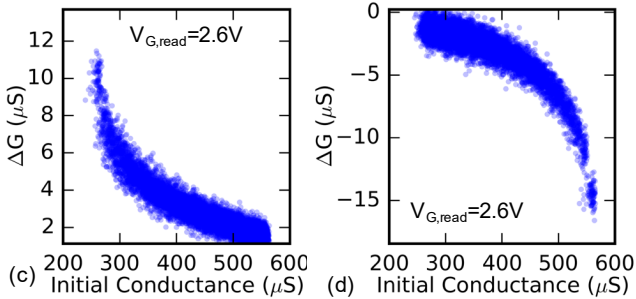

Fig 11: At $V_{G,read}$=2.6V the conductance change in asymmetric between program and erase leading to lower training accuracy.

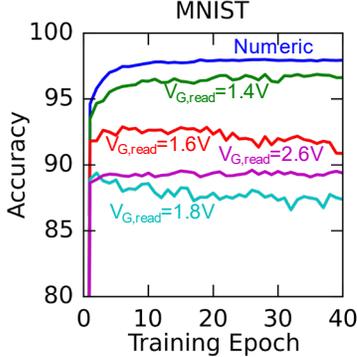

Fig 12: The lower the read voltage, the higher the training accuracy. A single device is used per weight.

TABLE I
A/D AND D/A CONVERTER PROPERTIES

|  | Range | Bits |
|---|---|---|
| **Row Input** | -1 to 1 | 8 |
| **Col Output** | -6 to 6 | 8 |
| **Col Input** | -1 to 1 | 8 |
| **Row Output** | -4 to 4 | 8 |
| **Row Update** | -0.01 to 0.01 | 7 |
| **Col Update** | -1 to 1 | 5 |

TABLE II
DATASET PROPERTIES

| Data set | #Training/Test Examples | Network Size |
|---|---|---|
| File Types[7] | 4,501 / 900 | 256×512×9 |
| MNIST [8] | 60,000 /10,000 | 784×300×10 |

TABLE III
AREA COMPARISONS

|  | 8 bit | 4 bit | 2 bit |
|---|---|---|---|
| SRAM (μm$^2$) | 836,000 | 814,000 | 800,000 |
| ReRAM (μm$^2$) | 75,000 | 46,000 | 41,000 |
| SONOS (μm$^2$) | 195,000 | 166,000 | 161,000 |

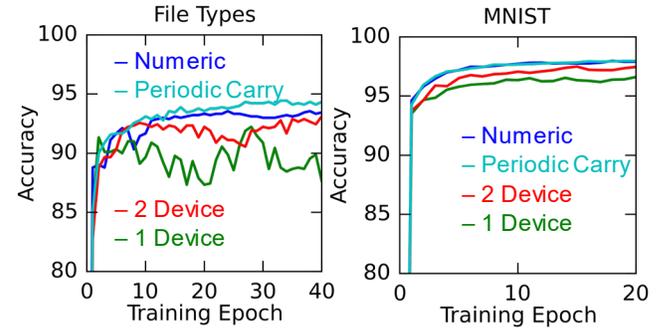

Fig 13: Training a neural network with the SONOS device can reach good accuracies of around 96% on MNIST when using a single device, but can reach ideal accuracies when using multiple devices with periodic carry. The 1 device architecture uses a single device to represent a weight and subtracts a reference current. The two device architecture takes the difference between two devices to represent negative numbers. The periodic carry architecture also uses two devices for the file types dataset and four devices for MNIST.

SONOS device a detailed system simulation was performed in CrossSim[3, 7], Sandia's analog crossbar simulator. We model the general purpose neuromorphic system in [3] where crossbars are used to perform matrix operations in analog and the inputs and outputs are processed in digital. This requires digital to analog (D/A) and analog to digital (A/D) converters at the inputs and outputs as specified in Table I. The bit precision and algorithmic input/output ranges used are given. They have a negligible (0.2%) impact on accuracy[5]. In order to model negative weights a single device per weight is initially used and reference current is subtracted[3]. Two different two-layer neural networks, summarized in Table II, are simulated [11, 12]. Simulation details are explained in the supplementary information of [7]. It's assumed that write voltages or pulse lengths can be scaled to vary the amount written.

As seen in Fig 12, by choosing the correct gate voltage, a good accuracy of 96.9% is achievable on MNIST. Representing negative numbers by taking the difference between two devices averages out some of the noise and increases the accuracy to 97.6% on MNIST. Using two devices per digit to represent negative numbers and two digits to represent a weight with periodic carry[5] an ideal device accuracy of 98.0% can be achieved as shown in Fig 13. We use a base 8, 2-digit number system where the first digit represents numbers 8 times larger than the second digit. Periodic carry allows one to take advantage of both a parallel write and a place value number system. Normally, a carry must be computed after every addition if using multiple digits. This eliminates the benefit of the parallel update. Allowing for part of an analog device's conductance range to represent a carry allows the carry from the second digit to the first digit to be computed only once every 1000 updates, thereby averaging out the cost of reading each memory element and adjusting the weights to perform a carry. We dedicate 50% of the conductance range of the lowest order digit to representing the carry.

For the file types dataset, only a single device is needed per digit and using periodic carry actually results in a higher accuracy than the numeric floating-point calculation (likely due to noise finding a more optimal solution).

TABLE IV
ENERGY AND LATENCY COMPARISONS

|  | VMM | | | MVM | | | OPU | | | Total | | |
| --- | --- | --- | --- | --- | --- | --- | --- | --- | --- | --- | --- | --- |
|  | 8 bit | 4 bit | 2 bit | 8 bit | 4 bit | 2 bit | 8 bit | 4 bit | 2 bit | 8 bit | 4 bit | 2 bit |
| Energy – SRAM (nJ) | 2850 | 2237 | 1848 | 4855 | 4241 | 3852 | 4300 | 3673 | 3274 | 12,000 | 10,150 | 8974 |
| Energy – ReRAM (nJ) | 12.8 | 1.00 | 0.44 | 12.8 | 1.00 | 0.44 | 2.2 | 1.00 | 0.46 | 27.9 | 2.66 | 1.35 |
| Energy – SONOS (nJ) | 14.4 | 2.25 | 1.5 | 14.4 | 2.25 | 1.5 | 71.5 | 30.9 | 10.6 | 100 | 35.4 | 13.6 |
| Latency – SRAM (μs) | 4 | 4 | 4 | 32 | 32 | 32 | 8 | 8 | 8 | 44 | 44 | 44 |
| Latency – ReRAM (μs) | 0.384 | 0.024 | 0.011 | 0.384 | 0.024 | 0.011 | 0.512 | 0.032 | 0.032 | 1.28 | 0.080 | 0.054 |
| Latency – SONOS (μs) | 0.402 | 0.032 | 0.014 | 0.402 | 0.032 | 0.014 | 20 | 20 | 20 | 20.80 | 20.06 | 20.02 |

## IV. ARCHITECTURAL EVALUATION

One of the key drawbacks of using a floating gate memory for an analog accelerator is that it requires a far larger area and voltage versus a ReRAM. Nevertheless, it is still possible to achieve significant system level advantages relative to an optimized digital SRAM based ASIC. To understand this, the architectural level analysis in [1] was modified to use a 1024x1024 SONOS array. The energy, area, and latency of a neural core that performs the three key matrix operations, VMM, MVM, and OPU was modelled. A 14/16nm process was modelled for the digital logic and interconnects. We assume the SONOS cell can scale to 28 nm and estimate a gate capacitance of 100aF and cell area of 0.053μm$^2$ based on existing 28nm floating gate transistors[13, 14]. We also assume it's possible to optimize the channel to give the high resistance (100 MΩ) needed for large scale arrays. The access transistors are assumed to have the same area and capacitance as the floating gate cell. Lastly, writing the array requires large high voltage transistors that can support 11V. Based on [15], high voltage vertical transistors can be fabricated in an area of 1.44 μm$^2$ and capacitance of 7.44 fF. These transistors are 9% of the core area. If needed, larger planar high voltage transistors can be used without drastically changing the overall area. We assume a future process will be able to integrate the needed transistors on a single substrate as commercial 28nm embedded flash is already in development. The ReRAM and SRAM based accelerators and device properties are described in detail in [1]. The SRAM based accelerator is based on A 1MB cache synthesized using a cache generator targeting the 14/16 nm PDK. The ReRAM is assumed to have a 100 MΩ on state, 35 aF capacitance, 10X on/off ratio and a 1.8V write voltage. The resulting energy, area, and latency relative to Digital SRAM and Analog ReRAM based accelerators is summarized in Tables III & IV for the accelerator. For an eight-bit floating gate training accelerator, 70% of the write energy is due to the CV$^2$ energy of charging wires to 10 or 11V. The very low write currents result in negligible contributions to the write energy. The SONOS read latency is comparable to ReRAM as the timing is dominated by the A/D and D/A converters. However, 96% of the total latency is due to the slow write speed of SONOS. Nevertheless, the large parallelism afforded by an analog accelerator allows for the total SONOS latency to still be 2X faster than an SRAM based accelerator. Latency can be decreased by trading off retention for a faster write or by using a device with a steeper subthreshold swing that allows for a larger conductance change with a smaller threshold shift.

Only 57% of the area is due to the SONOS cell and access transistor, indicating that the array area is reasonably balanced with the area of the rest of the circuitry. If higher area efficiency is desired two 3d integration options can be explored. High density (1.8μm pitch) face to face interconnects[16] could be used to connect two wafers, one with digital logic, and one with high voltage and floating gate transistors to reduce the area by 50%. The 3d interconnect capacitance would be less than the row or column capacitance in the SONOS array. Following [17], 3D NAND arrays could also be used to store multiple layers of a neural network in the same 2D area. Each individual SONOS cell in Fig 1 could be replaced by a column in a 3D NAND array.

## V. CONCLUSION

Floating gate memories, currently available in commercial foundries, are a compelling near-term option for analog training accelerators. This work has demonstrated lower write noise and write nonlinearity than alternative resistive memories, allowing for training to ideal accuracies on MNIST. Despite the high voltage and slow writes, the energy, area, and latency of an 8 bit floating gate neural accelerator is still 120X, 5.0X and 2.1X better respectively than an optimized digital ASIC counterpart. The high accuracies are enabled by operating the devices in the subthreshold regime giving symmetric write nonlinearities. Any three-terminal transistor based device should be able to operate in this favorable regime.